\documentclass{ws-ijmpa}
\usepackage{graphicx}

\begin{document}

\markboth{B.M.K. Nefkens}{Crystal Ball Highlights}

\catchline{}{}{}{}{}

\title{Highlights of Crystal Ball Physics}

\author{B.M.K. Nefkens \emph{for the Crystal Ball Collaboration}}
\address{UCLA, Los Angeles, CA 90095, USA}


\maketitle


\begin{abstract}
Differential and total cross sections are presented for $\pi^-$- and
$K^-$-induced reactions on a proton  target leading to all-neutral
final states.  Also shown are rates for rare and upper  limits for
forbidden eta-meson decays to test Chiral Perturbation Theory, the
$\pi^0\pi^0$ interactions and $C$ and $CP$ invariance. 
\end{abstract}



\section{Introduction}

The Crystal Ball (CB) is a multiphoton spectrometer whose outstanding
features include an acceptance that is 93\% of 4$\pi$ steradian; it
uses a simple trigger which is based on the total energy deposited in
the ball. The CB typically measures seven reactions simultaneously
covering the full $180^{\circ}$ angular range at once.  These features
have made the CB an exceptionally productive detector.  The main
detection elements are 672 isolated NaI(Tl) crystals arranged in two
matching hemispheres with 336 windows each.  There is an entrance and
exit channel for the beam, and a cavity in the center for a liquid
$\mathrm{H}_2$ target.  A veto-barrel counter surrounds the target.
The CB has excellent energy and spatial
resolutions\cite{Starostin,Prakhov1}.  The CB can detect many photons
at once, making it our detector of choice for measuring the production
rates and spectra of final states which include $\gamma$, $\pi^0$,
$2\pi^0$, $3\pi^0$, and $\eta$.  We have demonstrated that the CB is
also an excellent detector for neutrons as well as for
$\Lambda\to\pi^0n\to2\gamma n$,
$\Sigma^0\to\gamma\Lambda\to\gamma\pi^0n\to3\gamma n$,
$\Sigma^+\to\pi^0p\to2\gamma p$, and $K_s\to2\pi^0\to4\gamma$. 

We have performed four types of experiments using the AGS C6 beam line
with separated $\pi^-$ and $K^-$ beams.  The maximum beam momentum was 750
MeV/$c$. The CB is now at Mainz for an extensive program of linearly and
circularly polarized photons interacting with a polarized proton or
deuteron target producing the same particles that we measured at the
AGS, with $E_\gamma$ up to 1.5 GeV.  We also have a program on
$\eta$, $\eta'$, and $\omega$ decays.  The priority measurements are
those of the magnetic dipole moment of the $\Delta^+(1232)$ and
$N^+(1535)$ resonances.  Other reports on AGS work with the CB include
inverse photoproduction, shown by W. Briscoe; $\pi^-p$ charge exchange
(CEX) in the region of the $\eta$ cusp, shown by S. Starostin; and
low-energy CEX, shown by M. Sadler.

The CB is especially suited for intermediate energy physics, which
includes: 
\begin{description}
\item[a)]hadron structure and the determination of the mass, width,
  decay branching ratios, and magnetic dipole moment of baryonic
  resonances; 
\item[b)]hadron-hadron interactions to obtain meson-meson and
  meson-nucleon scattering lengths and isospin and SU(3) mixing
  angles; 
\item[c)]testing the applicability of effective Lagrangian models such
  as Chiral Perturbation Theory, $\chi PTh$, the $1/N_c$ expansion,
  etc.; 
\item[d)]determine the ratio of and the difference between the light
  quark masses; 
\item[e)]investigations into the broken symmetries: flavor, isospin,
  $G$-parity, chiral symmetry, $C$, $CP$, $T$, and $CPT$ invariance;
\item[f)]searches for new particles: leptoquarks, supersymmetric
  particles, and pentaquarks; 
\item[g)]investigations of medium modifications using $f_0(600)$,
  $\omega$, and $\rho$ production in nuclei, and the dependence of
  isospin breaking on the nuclear environment. 
\end{description}

We have already published the most interesting results in 5 PRL's and 
9 PR's as well as in 4 papers in
other refereed journals. It facilitates our giving the highlights of 
the CB program at the AGS.

\section{Eta-Meson Production by $\pi^-$ and $K^-$}

The production of the $\eta$ in the reactions $\pi^-p\to\eta n$ and
$K^-p\to\eta\Lambda$ from threshold to 750 MeV/$c$ has been measured
in detail\cite{Starostin}.  When comparing the results for the two
reactions one is struck by the similarities between them:
\begin{description}
\item[a)]the identical, sharp onset of $\eta$ production directly at
  the opening of the $\eta$ channel;
\item[b)]the total cross section $\sigma_t$ has a strong peak not far
  above threshold for both reactions; 
\item[c)]for threshold S-wave production, described by
  $\sigma_t=C\tilde{p}_\eta$, the constant $C$ is the same for both
  reactions within errors; 
\item[d)]the angular distributions of the differential cross sections
  have a similar shape, which is like a shallow bowl, for both
  reactions; 
\item[e)]the $\eta n$ and $\eta\Lambda$ scattering lengths are both
  large and attractive; 
\item[f)]the intermediate states, the $N(1535)$ and the
  $\Lambda(1670)$ have the same spin and parity ($J^P=\frac{1}{2}^-$)
  and the same SU(3) classification; 
\item[g)]the branching ratio into the $\eta$ channel of both
  intermediate states is anomalously large. 
\end{description}
The above is readily explained by invoking dynamic flavor symmetry.


\section{Dynamic Flavor Symmetry}

According to QCD, the strong interaction of all quarks is identical in
the limit of zero mass quarks.  When the baryons are three-quark and
mesons quark-antiquark states, the relevant symmetry is SU(3)$_F$.  The
$\Lambda$ and $n$ are then degenerate, as are $\pi$ and $K$ as well as
the $N(1535)$ and $\Lambda(1670)$, which are the intermediate states
in $\eta$ production by $\pi^-$ and $K^-$.  Consequently, in the limit
of massless quarks, $\eta$ production by $\pi^-$ and $K^-$ should be
identical.  When the quarks have mass the QCD Lagrangiain must be
expanded by the quark-mass term, ${\cal L}_m$, where ${\cal
  L}_m=\overline{q}m_qq$.  The effect of the mass term is to break the
degeneracy.  This causes the physical particles to have different
masses, which leads to a different phase space for various processes,
but it should not change much the overall dynamics of a reaction.
There should be no radically different new features for two formerly
degenerate SU(3)$_F$ symmetric reactions.  The gross dynamic features
such as a sharp onset of $\eta$ production, the $S$-wave dependence of
$\sigma_t$, and the shape of the angular distribution should be similar
for $\pi^- p\to\eta n$ and $K^- p\to\eta\Lambda$, as we have
experimentally observed and discussed\cite{Nefkens1}.

The consequences of the breaking of the $\mathrm{SU}(3)_F$ symmetry
due to the different masses of the quarks depends on the parameter and
the process.  SU(3)$_F$ implies that there is one $\Lambda^*$
resonance of the same spin and parity for every $N^*$ resonance, which
is observed experimentally.  The breaking due to the mass term causes
the $\Lambda^*$'s to be some 140 MeV heavier than their $N^*$
partners, as is also observed.  The ${\cal L}_m$ term lies at the
heart of the success of the famous Gell-Mann-Okubo mass relations
between the ground states of the octet and decuplet baryons.

The importance of SU(3)$_F$, even when it is broken, is the role it
can play in the hunt for exotica, the non-three-quark baryons, the
hybrids, the meson-baryon bound states and the non-unique pentaquarks.
SU(3)$_F$ provides a novel way to search indirectly for the so called
missing $N^*$ and $\Lambda^*$ resonances.  SU(3)$_F$ implies an
important relation between the baryons: for every $N^*$ state there
must be the three partners, a $\Lambda^*$, $\Sigma^*$, and $\Xi^*$
octet state with the same spin and parity and heavier by one or two
$\Delta M_{sd}$ which is the mass difference between the $s$ and $d$
quarks.  For every $\Delta^*$ there must be three decuplet partners, a
$\Sigma^*$, $\Xi^*$, and $\Omega^*$ hyperon with the same $J^P$ and
one, two, or three $\Delta M_{sd}$ heavier.  There are also a few
$\Lambda^*$ SU(3) singlet states.  A hyperon that does not fit this
recipe is exotic, or a missing hyperon.  Some examples:
\begin{description}
\item[a)]The $\Sigma$(1580)$\frac{3}{2}^-$ is only a one-star state.
  There is no missing or matching $N^*$ or $\Delta^*$ in the relevant
  mass region.  Recent measurements by the CB of $K^-p\to\pi^0\Lambda$
  have not found evidence for this state\cite{Olmsted}.  Very likely
  this $\Sigma$(1580) does not exist and should be removed from the
  particle-properties compilation\cite{Eidelman}. 
\item[b)]The $\Xi$(1620) is another one-star state awaiting
  confirmation (or removal); the mass is too low to be a $3q$ state. 
\item[c)]SU(3)$_F$ implies that the $\Xi$(1690) has
  $J^P=\frac{1}{2}^+$.  The second lightest excited state of the
  nucleon is the Roper with $J^P$ = $\frac{1}{2}^+$ so the second
  lightest $\Xi^*$ should be the Roper analog and have $J^P =
  \frac{1}{2}^+$ also.
\end{description}

One of the reasons it has been so difficult to investigate the
so-called missing $N^*$ and $\Delta^*$ resonances is the fact that
these states are expected to be very broad with $\Gamma$ of 400--500
MeV.  They overlap with each other and with neighboring states.  It so
happens that all known $\Xi^*$ resonances are narrow.  A careful
search for new cascade states could confirm that certain ``missing''
$N^*$ or $\Delta^*$ are also missing $\Xi$ states.  This would
indicate that they actually are not missing states but do not exist. 

A subgroup of SU(3)$_F$ is SU(2)$_F$.  It is the basis of isospin,
charge symmetry and G-parity invariance.  Since the up-down quark mass
difference is only a few MeV, there is abundant evidence for
approximate isospin invariance in nuclear physics such as the near
equality of the masses of isospin multiplet partners.  There is also
ample evidence for dynamic SU(2) flavor symmetry in the near equality
of many isospin related reaction cross sections and angular
distributions. 

\section{$\pi^0\pi^0$ Production by $\pi^-$ and $K^-$}

The Crystal Ball is eminently suited for measuring $\pi^0\pi^0$
production by $\pi^-$, $K^-$, and $\gamma$.  It will be interesting to
see whether dynamic flavor symmetry is appplicable to three-body final
state reactions with their greater complexity.  Consider the case of
$K^- p\to\pi^0\pi^0\Lambda$ at $p_K$ = 750 MeV/$c$.  We can assume
that the reaction is dominated by the s-channel processes.  There are
two quite different processes which are likely to be important:
\begin{description}
\item[a)]$f_0$(600) production from the two-body decay of the
  $\Lambda^*$ intermediate state, followed by the decay of the $f_0$
  into $\pi^0\pi^0$, thus, $K^-p\to\Lambda^*\to
  f_0\Lambda\to\pi^0\pi^0\Lambda$.  The dominant $\Lambda^*$
  intermediate states are the $\Lambda(1670)\frac{1}{2}^-$ and/or the
  $\Lambda(1690)\frac{3}{2}^-$.  This process is characterized by a
  broad, uniform band in the Dalitz Plot (DP) centered around the mass
  of the $f_0$(600).  In our choice of coordinates, which is
  $\tilde{m}^2(\pi^0\pi^0)$ for the vertical axis where
  $\tilde{m}(\pi^0\pi^0)$ is the invariant mass of the $\pi^0\pi^0$
  system, while the horizontal axis shows $\tilde{m}^2(\pi^0\Lambda)$,
  the $f_0$ band should be horizontal. 
\item[b)]The competing possibility is the sequential decay of two
  hyperons, specifically,
  $K^-p \to \Lambda^* \to \pi^0\Sigma(1385)\frac{3}{2}^+ \to
  \pi^0\pi^0\Lambda$.  This mode is characterized by a strong vertical
  band with $\tilde{m}^2(\pi^0\Lambda)$ centered around the mass of
  the $\Sigma^0(1385)$. 
\end{description}

There are two indistinguishable $\pi^0$'s in the final state.  We
don't know which $\pi^0$ comes from $\Lambda^*$ and which from
$\Sigma^0$ decay.  We solve this dilemma by plotting both options on
the DP.  This has as a consequence that the vertical band will be
slightly slanted.  Furthermore, there are two $\pi^0\Lambda$
scattering amplitudes contributing to the final state which interfere
with one another, causing the vertical band to be non-uniform.
Destructive interference could even cause a void in the band.  When
option a) is dominant we expect that the related reaction $K^-p \to
\pi^0\pi^0\Sigma^0$ will have a $\sigma_t$ similar to $K^-p \to
\pi^0\pi^0\Lambda$, and that the two reactions will have similar DP's.
For option b) the intermediate state to the $\pi^0\pi^0\Sigma^0$ final
state are $\Lambda^*$ resonances with different $J^P$, and we expect
different $\sigma_t$'s and DPs.  Experimentally at 750 MeV/$c$ we find
$\sigma_t(K^- p \to \pi^0\pi^0\Lambda = 6\sigma_t(K^-p \to
\pi^0\pi^0\Sigma^0)$ and the DPs are different\cite{Prakhov2,Prakhov3}
and there is no uniform horizontal band, thus the sequential decay of
2 hyperon resonances plays the dominant role in 2$\pi^0$ production. 

Flavor symmetry relates $\pi^0\pi^0$ production by $K^-$ to that by
$\pi^-$.  Around 750 MeV/c beam momentum the dominant channels are:
\begin{description}
\item[A)]$K^-p \to \Lambda^*\frac{3}{2}^- \to
  \pi^0\Sigma^0(1385)\frac{3}{2}^+ \to \pi^0\pi^0\Lambda$
\item[B)]$K^-p \to \Sigma^*\frac{3}{2}^- \to
  \pi^0\Lambda(1405)\frac{1}{2}^- \to \pi^0\pi^0\Sigma^0$
\item[C)]$\pi^-p \to N^*\frac{3}{2}^- \to
  \pi^0\Delta(1232)\frac{3}{2}^+ \to \pi^0\pi^0n$
\item[D)]$\gamma p \to N^*\frac{3}{2}^- \to
  \pi^0\Delta(1232)\frac{3}{2}^+ \to \pi^0\pi^0 n$
\end{description}
Reaction D) is $\pi^0\pi^0$ photoproduction which is an
electromagnetic interaction and QCD does not apply to the $\gamma p$
system.  In reaction C) and D) the intermediate $N^*$'s actually are
different mixtures of the $N^*\frac{3}{2}^-$ and the
$N^*\frac{1}{2}^-$.  We expect the DP for C) and D) not to be the
same. 

Every particle that appears in process A) is an SU(3)$_F$ analog state
of every particle in the same role as process C).  It follows then
that the DP for $\pi^0\pi^0\Lambda$ should have the same features as
$\pi^0\pi^0n$.  This is what is observed\cite{Prakhov1,Prakhov2}. 

On the other hand, the intermediate states in A) and B) are not flavor
analog states.  The $\Sigma(1385)\frac{3}{2}^+$ belongs to a decuplet,
as does the $\Delta (1232)\frac{3}{2}^+$.  The
$\Lambda(1405)\frac{1}{2}^-$ is a singlet state, thus we expect the DP
for B) to be different than for A) which is also
observed\cite{Prakhov2,Prakhov3}. 

\begin{figure}
\centerline{\includegraphics[width=5.5in]{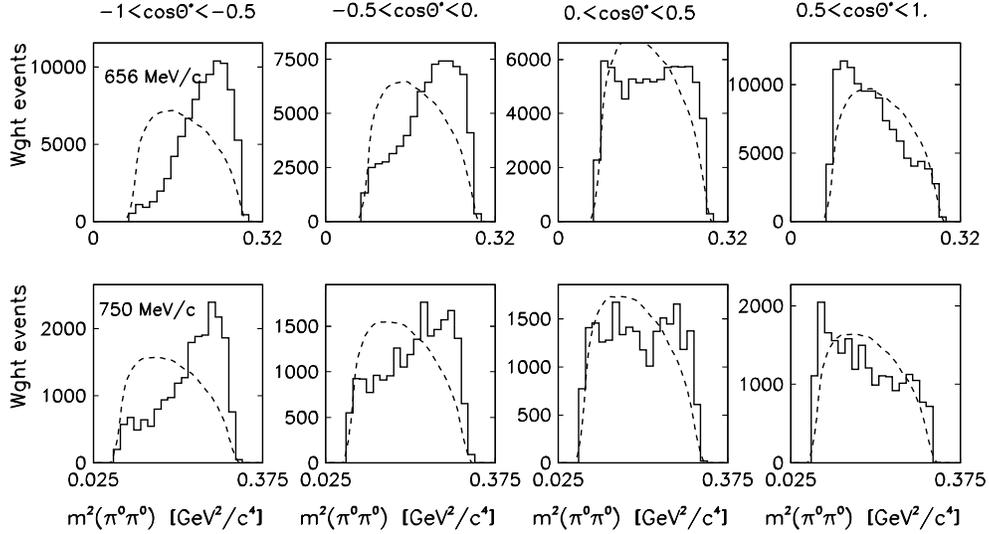}}
\caption{\label{fig}Top Row: $\pi^0\pi^0$ invariant mass spectra for
  $\pi^-p\to\pi^0\pi^0n$ at four angular intervals of
  $\theta_{\pi\pi}$.  The dashed line is the phase space
  distribution. Bottom Row: same as the top row, for the process
  $K^-p\to\pi^0\pi^0\Lambda$.} 
\end{figure}

We have sufficient $\pi^0\pi^0$ data to enable us to study the
dependence of the DP on its kinematic variables, specifically on
$\theta_{\pi\pi}$ which is the angle of the $\pi^0\pi^0$ system in the
center of mass and which is the complement of the recoil neutron angle
$\theta_n$. The DP of the 750 MeV/$c$ data depends strongly on
$\theta_{\pi\pi}$.  This is illustrated using the DP projection onto
the $\pi^0\pi^0$ invariant mass coordinate in Fig.~\ref{fig}. 
For the process $\pi^-p \to \pi^0\pi^0n$ at forward $\theta_{\pi\pi}$
the DP is peaked at small ${\tilde{m}}(\pi^0\pi^0)$, while for the
backward $\theta_{\pi\pi}$ the DP peaks at the highest available
$\tilde{m}(\pi^0\pi^0)$.  Dynamical SU(3) symmetry implies that the
$\theta_{\pi\pi}$ dependence is the same for A and C.  This is
observed, as seen in Fig.~\ref{fig}. In some sense the similarity in
the features of the four DP's for $\pi^0 \pi^0 \Lambda$ and $\pi^0
\pi^0 n$ in Fig.~\ref{fig} is one of the most spectacular
manifestations of flavor symmetry seen so far. 

There is much more interesting physics in the $\pi^0\pi^0$ production
data; space limitation prevents us from discussing it here.  We only
give two particularly striking examples. 
\begin{enumerate}

\item The excitations function of $\sigma_t(\pi^-p \to \pi^0\pi^0n)$
  has a clear shoulder at the pole position of the Roper (1440)
  resonance\cite{Prakhov1}.  It is a rare example of a direct
  manifestation of one of the most elusive states of the $N^*$ family.

\item Our data on $K^-p \to \pi^0\pi^0\Sigma^0$ at 750
  MeV/$c$\cite{Prakhov3} contains a good size sample of $\Lambda(1405)
  \to \pi^0\Sigma^0$ decays.  The spectral shape of this decay is of
  interest to many theorists who have models for the $\Lambda(1405)$
  being either a bound state or an overlapping set of two nearby
  states.  The data is open to discussion!
\end{enumerate}

\section{Eta-Meson Decays}

The decay mode $\eta \to 3 \pi^0$ provides a special test of our
understanding of the low-energy $\pi^0 \pi^0$ interaction, which is
the simplest known strong interaction and is free of the complication
associated with the Coulomb interaction of charged particle
processes.  The $\eta\to3\pi^0$ Dalitz plot has a three-fold symmetry
because of the three indistinguishable $\pi^0$'s, but it should not be
entirely uniform because the $\pi$-$\pi$ interaction is energy
dependent.  One can readily symmetrize the DP using as the variable
the distance z to the center of the DP\cite{Tippens}.  The projection
of the DP density along the z-axis is then given by $\vert A\vert =
1+\alpha z$,  where $\alpha$ is the slope parameter.  Four previous
experiments have been inconclusive.   Using a sample of 19 M $\eta's$,
the slope parameter was found to have the same sign (negative) as the
theoretical calculation by Kambor \emph{et al.}\cite{Kambor}, but
numerically two and a half times as big\cite{Tippens,Kambor}.

The decay width of the rare decay $\eta \to \pi^0\gamma\gamma$
provides a gold-plated test of chiral perturbation theory, $\chi PTh$,
because it is a direct test of the 3rd-order term.  This unique
feature comes about because the first order term is zero since the
neutral $\pi^0$ and $\eta$ meson do not couple in lowest order to the
photon.  The second order is ignorably small because the relevant
decay channels violate G-parity.  Various theoretical evaluations of
the decay rate with $\chi PTh$ or by other means such as vector-meson
dominance agree that $\Gamma(\eta \to \pi^0\gamma\gamma) \simeq
0.4$eV.  After 13 unconvincing attempts, there is only one modern
measurement.  It was made by GAMS-2000, who reported a $\Gamma$ of
about 0.8 eV, double the theoretical value.  The CB, using a sample of
28 M $\eta$'s, found 1600 events leading to $\Gamma(\eta \to
\pi^0\gamma\gamma) = 0.45\pm0.12$eV, a result that thrills the
theorists\cite{Prakhov4,Nefkens2}. 

There is considerable interest in improved tests of $C$-invariance of
the strong interactions.  Firstly, because of the observation of the
excess of matter over antimatter in the universe, it does not agree
with the big bang model.  Secondly, because of the structural
asymmetry of the Standard Model that groups the basic quark and lepton
constituents in left-handed doublets but right-handed singlets.

The $\eta$ meson makes possible several novel tests of $C$-invariance
by searching for $C$-forbidden decays into $\pi^0\pi^0\gamma$ and
$\pi^0\pi^0\pi^0\gamma$.  The sensitivity of these tests is readily
evaluated by comparison with $C$-allowed decays\cite{Nefkens2}.  The
CB has determined the first upper limit, $BR(\eta \to
\pi^0\pi^0\gamma) < 5\times10^{-4}.$  This implies that
$A^s_{\not{c}}/A_c < 2.5\times10^{-3}$ where $A^s_{\not{c}}$ is the
\emph{amplitude} of a possible $C$-violating isoscalar interaction and
$A_c$ is the allowed one.  The CB also found $BR (\eta \to \pi^0
\pi^0\pi^0 \gamma)<6 \times 10^{-5}$ which implies $A^v_{\not{c}}/A_c
< 3.4 \times 10^{-3}$ where $A^u_d$ is the isovector $C$-violating
amplitude. 

\section{Summary and Conclusion}

A selection from the data obtained with the Crystal Ball multiphoton
spectrometer at the AGS has been presented.  The results are used to
investigate various broken symmetries, in particular, flavor, isospin,
charge, and chiral symmetry, and $C$-invariance.  We have made a
direct test of third-order $\chi PTh$ in $\eta \to \pi^0\gamma\gamma$
and of the $\pi^0\pi^0 $ interaction via the slope parameter in $\eta
\to 3\pi^0$ decay.


\section*{Acknowledgements}

This work was supported by US DOE, NSF, NSERC of Canada, the MIS, and
RFBR from Russia.  We thank SLAC for the loan of the Crystal Ball.  The
help of BNL and the AGS with the set-up is much appreciated


\end{document}